\documentclass[a4paper,11pt]{article}
\usepackage{pos}
\usepackage{comment}

\title{A solution for infinite variance problem of fermionic observables}
\author*[a]{Hyunwoo Oh}
\author[a,b]{Andrei Alexandru}
\author[a]{Paulo F. Bedaque}
\author[b]{Andrea Carosso}

\affiliation[a]{Department of Physics, University of Maryland,\\
College Park, MD 20742}
\affiliation[b]{Department of Physics, The George Washington University,\\
Washington, DC 20052}

\emailAdd{hyunwooh@umd.edu}
\emailAdd{aalexan@gwu.edu}
\emailAdd{bedaque@umd.edu}
\emailAdd{acarosso@gwu.edu}

\abstract{ % for conference
Fermionic Monte Carlo calculations with continuous auxiliary fields often encounter infinite variance problem from fermionic observables. This issue renders the estimation of observables unreliable, even with an infinite number of samples. In this work, we show that the infinite variance problem stems from the fermionic determinant. Also, we propose an approach to address this problem by employing a reweighting method that utilizes the distribution from an extra time-slice. Two strategies to compute the reweighting factor are explored: one involves truncating and analytically calculating the reweighting factor, while the other employs a secondary Monte Carlo estimation. With Hubbard model as a testbed, we demonstrate that utilizing the sub-Monte Carlo estimation, coupled with an unbiased estimator, offers a solution that effectively mitigates the infinite variance problem at a minimal additional cost.
}

\FullConference{The 40th International Symposium on Lattice Field Theory (Lattice 2023)\\
July 31st - August 4th, 2023\\
Fermi National Accelerator Laboratory\\}

\begin{document}
\maketitle

\section{Introduction}

Monte Carlo methods have been successful to study non-perturbative physics phenomena, ranging from nuclear physics to condensed matter physics. However, they still suffer from many issues: sign problem~\cite{PhysRevB.41.9301}, ergodicity problem~\cite{Wynen:2018ryx}, and infinite variance problem, which make the estimation of observables exponentially hard. Especially, infinite variance problem causes the estimation of observables to be impossible since variances are divergent with the progress of Monte Carlo samplings.

One way to deal with the diverging variance for fermionic observables is to employ discrete auxiliary fields~\cite{Yunus:2022wdr}. However, one wants to use continuous auxiliary fields to use hybrid Monte Carlo~\cite{DUANE1987216} for faster convergence. Also, sign problem has been studied widely and some methods require the complexification of the integration domain~\cite{Alexandru:2020wrj, Berger:2019odf}, which can only be implemented with continuous variable Monte Carlo calculations.

In this work, we review infinite variance problem from fermionic observables and discuss its solution while employing continuous auxiliary fields. Specifically, we use Hubbard model as a testbed to show that the extra time-slice reweighting with sub Monte-Carlo methods can remove the diverging variance without additional errors.

Hubbard model, which is a strong candidate for explaining high-temperature superconductors, consists of hopping term, local interaction term, and chemical potential term:
\begin{equation}
    \begin{aligned}
    H  = & - \kappa \sum_{\langle x,y \rangle}(\hat{\psi}^{\dagger}_{\uparrow,x}\hat{\psi}_{\uparrow,y}+\hat{\psi}^{\dagger}_{\downarrow,x}\hat{\psi}_{\downarrow,y}) 
    + U \sum_{x}(\hat{\psi}^{\dagger}_{\uparrow,x}\hat{\psi}_{\uparrow,x}-\frac{1}{2})(\hat{\psi}^{\dagger}_{\downarrow,x}\hat{\psi}_{\downarrow,x}-\frac{1}{2}) \\
    & - \mu \sum_{x}(\hat{\psi}^{\dagger}_{\uparrow,x}\hat{\psi}_{\uparrow,x}+\hat{\psi}^{\dagger}_{\downarrow,x} \hat{\psi}_{\downarrow,x}-1).
    \end{aligned}
\end{equation}
On bipartite lattices, one can use the particle-hole symmetry to rewrite the Hamiltonian. By redefining $\hat{\psi}_{1,x} \equiv \hat{\psi}_{\uparrow, x}$ and $\hat{\psi}_{2,x} \equiv (-1)^x \hat{\psi}_{\downarrow, x}$, one can find that
\begin{equation}
    H =  - \kappa \sum_{\langle x,y \rangle}(\hat{\psi}^{\dagger}_{1,x}\hat{\psi}_{1,y}+\hat{\psi}^{\dagger}_{2,x}\hat{\psi}_{2,y}) 
    + \frac{U}{2}\sum_{x}(\hat{\psi}^{\dagger}_{1,x}\hat{\psi}_{1,x}-\hat{\psi}^{\dagger}_{2,x}\hat{\psi}_{2,x})^2 \\
    - \mu\sum_{x}(\hat{\psi}^{\dagger}_{1,x}\hat{\psi}_{1,x}-\hat{\psi}^{\dagger}_{2,x}\hat{\psi}_{2,x}).
\end{equation}

To remove the fermionic variables for Monte Carlo calculations using path integral formulation, one can use a Hubbard-Stratonovich transformation. Since there are some advantages for using compact auxiliary fields when one utilizes the contour deformation method for ameliorating sign problem, we choose the compact continuous Hubbard-Stratonovich transformation. Then one can find that (Details can be found in~\cite{Alexandru:2022dlq}.)
\begin{equation} \label{Eq:Expectation}
    \langle \mathcal{O} \rangle=
    \frac{\int D\phi \; \mathcal{O}(\phi) \; {\rm e}^{-S_0(\phi)} \det M_1(\phi) \det M_2(\phi)} {\int D\phi \; {\rm e}^{-S_0(\phi)} \det M_1(\phi) \det M_2(\phi)},
\end{equation}
where
\begin{equation}
    S_0(\phi) = -\beta \sum_{x,t}  \cos\phi_{t,x} \; \text{and} \; M_a(\phi) = \mathbb{I} + B_a(\phi_N) \cdots B_a(\phi_1).
\end{equation}
Each term in the fermion matrices is written as
\begin{equation}
B_a(\phi_t) = {\rm e}^{- H_2} {\rm e}^{- \tilde H_4(\phi_t)},
\end{equation}
where
\begin{equation}
    \begin{aligned}
    (H_2)_{x,y} & = \kappa \epsilon \delta_{\langle x,y \rangle}  +  \varepsilon_a \mu \epsilon  \delta_{x,y}, \\ 
    \tilde H_4(\phi_t)_{x,y} & = -i \varepsilon_a \sin\phi_{t,x} \delta_{x,y}.
    \end{aligned}
\end{equation}
Here, $\delta_{\langle x,y \rangle}$ is the nearest-neighbor hopping matrix, and $\varepsilon_1=+1, \; \varepsilon_2=-1$. The parameter $\alpha$ is related to the potential $U$ by
\begin{equation}
{\rm e}^{-\epsilon U/2} = 
\frac{I_0(\sqrt{\alpha^2-1})}{I_0(\alpha)}.
\end{equation}

Since $M_2(\phi)=M_1(\phi)^*$ at the half-filling, i.e. $\mu=0$, Hubbard model does not have the sign problem. In this work, we will only consider the half-filling case to remove the effect of the sign problem. Also, we will use the unit $\kappa \equiv 1$.

\section{Infinite variance problem}

\begin{figure*}[t!]
\includegraphics[width=0.98\textwidth]{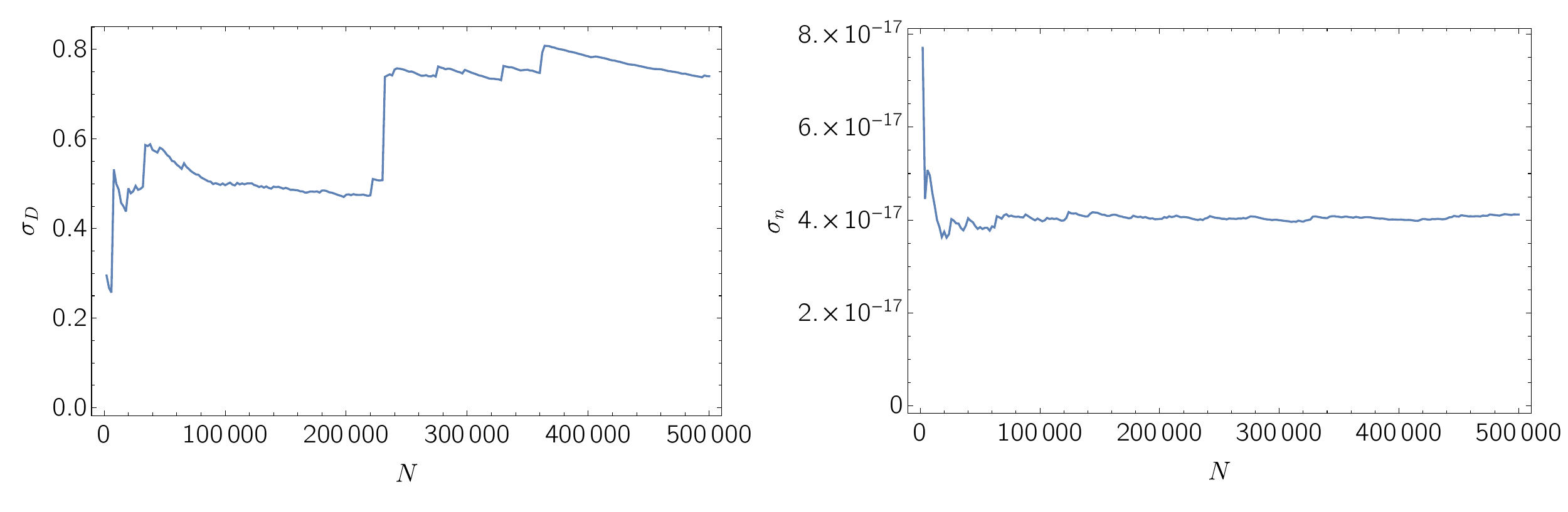}
\caption{Cumulative estimations of standard deviations using standard Monte Carlo calculations of Hubbard model. The left and right plot shows the standard deviations of double occupancy and density, respectively.}
\label{fig1}
\end{figure*}

Let us consider the expectation value of fermionic observables, i.e. $\mathcal{O} = f(\bar\psi, \psi)$. Using the Hubbard-Stratonovich transformation and the Gaussian Grassmann integration formula, one can find that
\begin{equation}
\langle \mathcal{O} \rangle=
\frac{\int D\bar{\psi} D\psi \; \mathcal{O} \; {\rm e}^{-S(\bar{\psi}, \psi)} }{\int D\bar{\psi} D\psi \; {\rm e}^{-S(\bar{\psi}, \psi)}}
= \frac{\int D\phi \; g\left( M_{ij}(\phi) \right) \; {\rm e}^{-S_0(\phi)}}{\int D\phi \; {\rm e}^{-S_0(\phi)} \det M(\phi)}, 
\label{Eq:O}
\end{equation}
where $g$ is a polynomial. Since the observable is proportional to the inverse of the fermion determinant, i.e., $g \propto 1/\det M(\phi)$. This holds for ``exceptional concifurations'' where $\det M=0$. Near the exceptional configurations where the determinant is nonzero but very small, the observable is very large, which makes variance jumps in the left panel of Fig.~\ref{fig1}. While the expectation value is finite when one does the integration of Eq.~(\ref{Eq:O}), the divergence can be infinite since the variance has the term proportional to the square of observable: $\sigma^2_\mathcal{O} = \langle \mathcal{O}^2 \rangle - \langle \mathcal{O} \rangle^2$. In terms of the path integral representation,
\begin{equation}
    \langle \mathcal{O}^2 \rangle = \frac{\int D\phi \; g^2 \left(M_{ij}(\phi) \right) \frac{1}{\det M(\phi)} \; {\rm e}^{-S_0(\phi)}}{\int D\phi \; {\rm e}^{-S_0(\phi)} \det M(\phi)}. \label{Eq:O^2}
\end{equation}
Therefore, since $g^2 {\rm e}^{-S_0} \geq 0 $ cannot remove the singularity from $1/\det M$ unless $g=0$ with the same order as the exceptional configurations, one cannot avoid the diverging variance if the determinant of fermion matrices have zero points.

There is an example in Hubbard model which shows that the infinite variance problem stems from the fermion determinant. Let us consider the two observables, double occupancy and density, in terms of fermion matrices:
\begin{equation}
    \begin{aligned}
        D(\phi) & = \frac{1}{V} \sum_x \langle n_{\uparrow}(x) n_{\downarrow}(x) \rangle = \frac{1}{V} \sum_x M^{-1}_2(\phi)_{x,x} \left(1 - M^{-1}_1(\phi)_{x,x} \right), \\
        n(\phi) & = \frac{1}{V} \sum_x \langle n_{\uparrow}(x) + n_{\downarrow}(x) \rangle = \frac{1}{V} \sum_x \left( M^{-1}_2(\phi)_{x,x} -  M^{-1}_1(\phi)_{x,x}\right).
    \end{aligned} \label{Eq:obs}   
\end{equation}
Fig.~\ref{fig1} exhibits the cumulative estimations of standard deviations for each observable. It shows that the double occupancy has infinite variance problem while the density does not. This is because the double occupancy has the term proportional to the inverse of fermion matrices, i.e., $M^{-1}_1 M^{-1}_2$ in Eq.~(\ref{Eq:obs}), while the density only has a part of them, i.e., $M^{-1}_1$ or $M^{-1}_2$.

\section{Extra time-slice}

In the previous section, it was shown that infinite variance problem of fermionic observables comes from the exceptional configurations. One possible solution is to use a different distribution for Monte Carlo samplings and employ the reweighting method. It was suggested in~\cite{Shi:2015lyu} that one can utilize the distribution from extra time-slice. Let us consider that our path integral representation in Eq.~(\ref{Eq:Expectation}) is trotterized with N time-slices. If one defines
\begin{equation}
    F(\phi) \equiv \int {\rm d}\phi^* \; {\rm e}^{-S_0(\phi^*)} \det M_{N+1}(\phi, \phi^*),
\end{equation}
where $M_{N+1}$ is the fermion matrix with $N+1$ time-slices, one can find the partition function as
\begin{equation}
        Z = \int [{\rm d}\phi]_N {\rm e}^{-S_0(\phi)} \det M_N(\phi)
\frac{F(\phi)}{F(\phi)} = \int [{\rm d}\phi]_N {\rm d}\phi^* \;  R(\phi) \; {\rm e}^{-S_0(\phi,\phi^*)}
\det M_{N+1}(\phi,\phi^*),
\end{equation}
where $[{\rm d} \phi]_N$ denotes the path integral measure with N time-slices. Then observables can be estimated with the conventional reweighting procedure:
\begin{equation}
    \langle \mathcal{O}\rangle_N = \frac{\langle \mathcal{O}(\phi)  R(\phi) \rangle_{N+1}}{\langle R(\phi) \rangle_{N+1}},
\end{equation}
where $R(\phi) = \det M_N(\phi) / F(\phi)$ and the subscript $N+1$ denotes that the Monte Carlo samples are chosen from the new distribution:
\begin{equation}
p_{N+1}(\phi,\phi^*) = 
\frac{{\rm e}^{-S_0(\phi,\phi^*)} \det M_{N+1}(\phi,\phi^*)} {\int[{\rm d}\phi]_N{\rm d}\phi^*
{\rm e}^{-S_0(\phi,\phi^*)} \det M_{N+1}(\phi,\phi^*)}.
\end{equation}
With this new distribution, infinite variance problem is cured since the variance involves $\mathcal{O} R$, which does not have any singularities. Therefore, the task is to estimate the reweighting factor $R(\phi)$.

\section{Unbiased estimator}

In~\cite{Shi:2015lyu}, the authors suggested that one can integrate $F(\phi)$ analytically using BSS formula~\cite{BSS} and expand it in $\epsilon \equiv \beta/N$:
\begin{equation}
    \begin{aligned}
    F(\phi) & \equiv \int {\rm d}\phi^* \; {\rm e}^{-S_0(\phi^*)} \det M_{N+1}(\phi, \phi^*) = \mathrm{Tr} \left[ \mathrm{e}^{- \tilde H_2} \mathrm{e}^{-\epsilon H_4} B(\phi_N) \cdots B(\phi_1) \right] \\
    & = \mathrm{Tr} \left[ (1 - \epsilon H ) B(\phi_N) \cdots B(\phi_1) \right] + O(\epsilon^2) 
    = (1-\epsilon H(\phi) )\det M_N(\phi) + O(\epsilon^2).
    \end{aligned} \label{Eq:analytic}
\end{equation}

The advantage of this method is that it does not have any additional cost except the increased cost from the new auxiliary field, but there are some disadvantages. First of all, there is a perturbative error from the truncation and the number of Wick contractions increases exponentially as one goes to higher orders. Also, one needs small $\epsilon$ since $F(\phi)$ can be zero, which can generate another infinite variance problem.

Instead of using analytical method, one can directly estimate $F(\phi)$ using Monte Carlo calculations using the new auxiliary field $\phi^*$ (which we call sub-MC method):
\begin{equation}
    F(\phi) = Z\langle \det M_{N+1}(\phi,\phi^*) \rangle_g,
\end{equation}
where the subscript $g$ represents the Monte Carlo samplings using ${\rm e}^{-S_0(\phi^*)}$.

However, what one needs to estimate is $1/F(\phi)$, not $F(\phi)$, and it can be easily checked that just taking an inverse of $F(\phi)$ is biased:
\begin{equation}
    \begin{aligned}
       \left\langle \frac{1}{\overline{A}} \right\rangle = \frac{1}{\langle  A\rangle}  -
       \left\langle \frac{\overline{A}-\langle { A} \rangle }{ \left\langle A \right\rangle ^2}
       \right\rangle
        +
        \left\langle \frac{ \left(\overline{ A}-\langle A \rangle \right)^2}{\langle A \rangle^3}
        \right\rangle
        - \cdots \,,
    \end{aligned} \label{Eq:biased}
\end{equation}
where $\overline{A}$ denotes the finite sample average of $A$. Note that the third term in Eq.~(\ref{Eq:biased}) is not zero.

Therefore, one needs to find an unbiased estimator for $1/F(\phi)$. In~\cite{Moka}, the authors suggested an unbiased estimator of $1/\langle A \rangle$:
\begin{equation}
    \hat \xi_{ A} \equiv \frac{w}{q_n} \prod_{i=1}^n (1-w { A}_i) \,.
\end{equation}
Here, $q_n$ is an arbitrary discrete probability distribution and $w < 1/\langle A \rangle$. Then the variance minimizing choice of $q_n$ and $w$ is 
\begin{equation}
    \begin{aligned}
        w & = {\rm min}\left\{ \frac{1}{k \overline{A}}, \frac{\overline{ A}}{\overline{ A^2}}, \frac{1}{A_{\rm max}} \right\}, \\
        p & = 1 - \left[1- 2 w \overline{ A} + w^2 \overline{ A^2} \right]^{\frac{1}{2}}, \\
        q_n & = p(1-p)^n,
    \end{aligned} \label{Eq:varmin}
\end{equation}
where $\overline{A}$ denotes the sample average of $A$.

\section{Result}

\begin{figure*}[t!]
\includegraphics[width=0.98\textwidth]{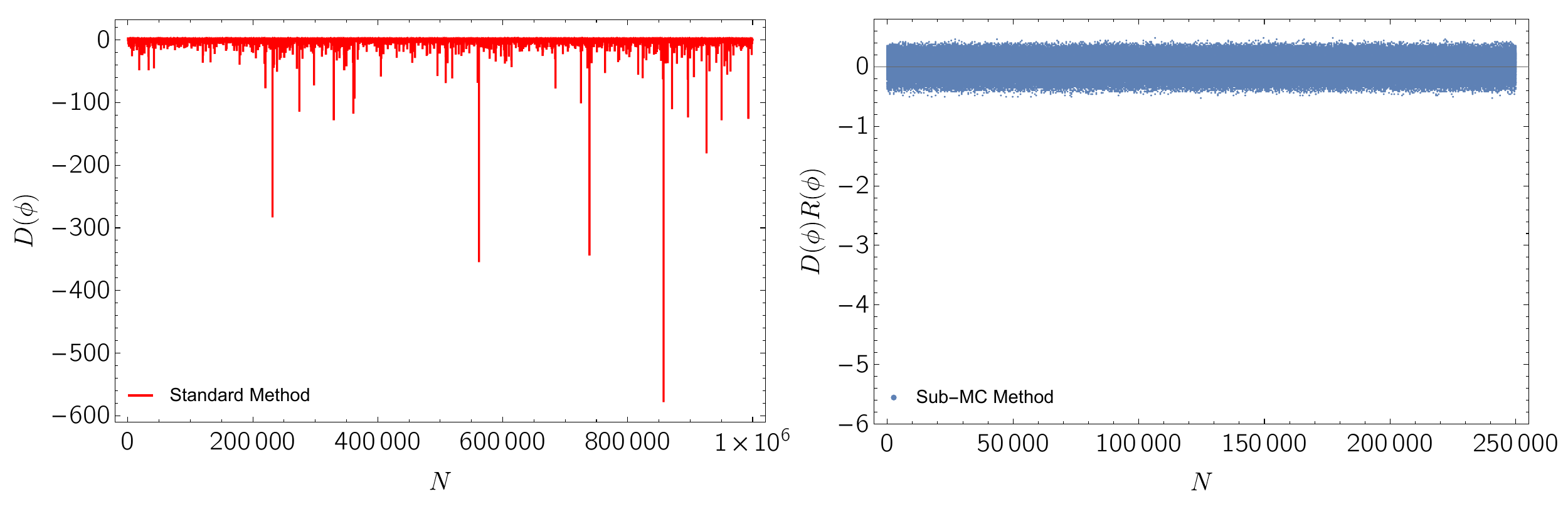}
\caption{Distributions of double occupancy on a $4\times4$ lattice, with $U=8$ and $\beta=2$. The left plot shows the standard Monte Carlo and the right plot exhibits the extra time-slice method with sub-MC sampling. Note that the two plots require different Monte Carlo runs since sub-MC method utilizes different probability distribution.}
\label{fig2}
\end{figure*}

Fig.~\ref{fig2} shows the distributions of double occupancy using the standard Monte Carlo and the extra time-slice reweighting with sub-MC method. The left panel has large negative peaks during the sampling, which makes the variance jumps in Fig.~\ref{fig1}. However, this abnormal behavior is well mitigated in the right panel, with the reweighting method. Note that the expectation value of double occupancy using the standard method is positive even though the exceptional configurations contribute to it negatively.

\begin{figure*}[t!]
\includegraphics[width=0.98\textwidth]{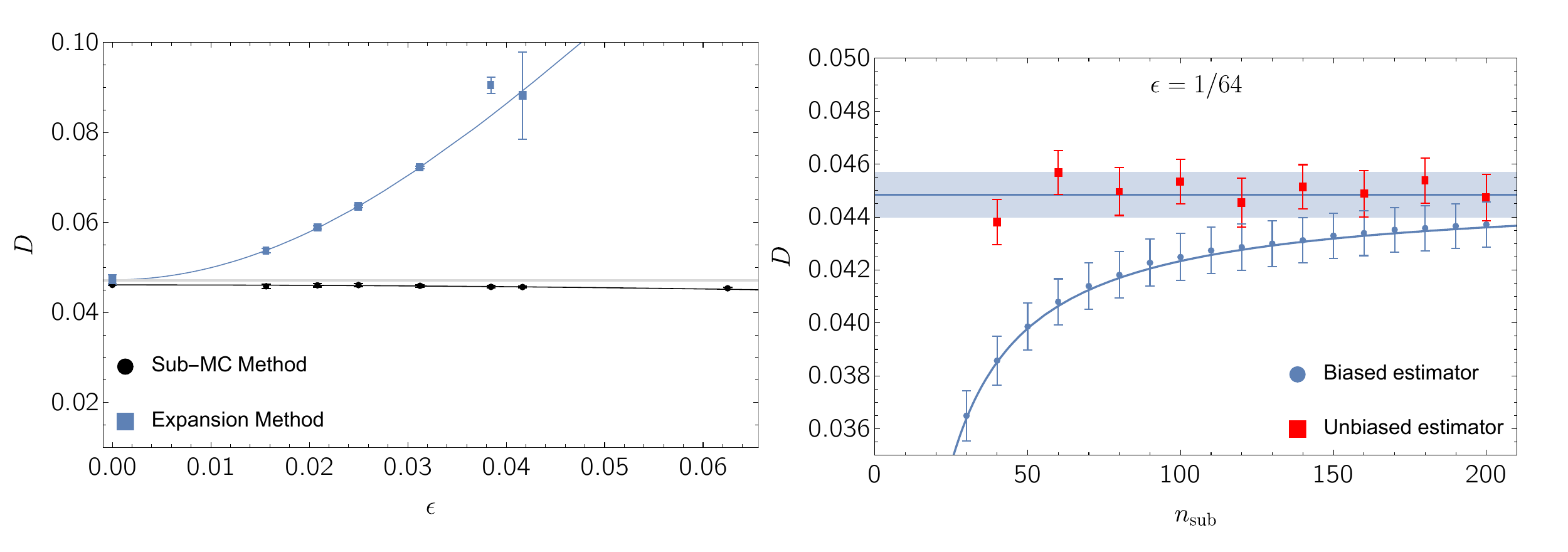}
\caption{Double occupancy using the analytic and sub-MC methods on a $4\times4$ lattice, with $U=8$, and $\beta=2$. The left plot compares the analytic and sub-MC methods. The gray band is the benchmark result in~\cite{Benchmark}. The right figure compares the biased and unbiased estimator. The solid line for the biased estimator is the $1/n_{\rm sub}$ fit. The blue band represents the $n_{\rm sub} \rightarrow \infty$ limit of the biased estimator with its error.
} \label{fig3}
\end{figure*}

The left panel of Fig.~\ref{fig3} exhibits the comparison of the analytic method in Eq.~(\ref{Eq:analytic}) and the sub-MC method. One can see that as the trotterization error increases, the Taylor expansion in terms of $\epsilon$ does not well behave, while the sub-MC method does not have the error from the truncation. Note that the fitting for the analytic method only used first four points because of increasing error at large $\epsilon$. It means that one needs to use small $\epsilon$ to employ the analytic method, and so the overall cost of Monte Carlo calculations can be cheaper for the sub-MC method even though it has the secondary Monte Carlo sampling procedure.

The right plot in Fig.~\ref{fig3} shows the the effect of the sub-MC method with the unbiased estimator. The bias in the biased estimator has  $1/n_\text{sub}$ behavior, where $n_{\rm sub}$ means the number of sub Monte Carlo samplings. The number of sub-MC samples for the unbiased estimator is randomly chosen from $q_n$ in Eq.~(\ref{Eq:varmin}) and therefore the average value $n_{\rm sub} \approx 2k$ is used for the figure, which includes the cost $k$ of the estimation for $w$ and $p$ in Eq.~(\ref{Eq:varmin}). The blue band is the $n_{\rm sub}\rightarrow \infty$ extrapolation of the biased estimator. It shows that the unbiased estimator converges to the unbiased value with lower cost.

\acknowledgments

This work was supported in part by the U.S. Department of Energy, Office of Nuclear Physics under Award Number(s) DE-SC0021143, and DE-FG02-93ER40762, and DE-FG02-95ER40907.

\end{document}